\documentstyle[prl,aps]{revtex}
\begin{document}
\draft

\title{High--Energy Multiparticle Distributions and a Generalized Lattice
Gas Model}

\author{A. D. Jackson$^{(1),(2)}$, T. Wettig$^{(1),(2)}$ and
        N. L. Balazs$^{(2)}$}

\address{$^{(1)}$NORDITA, Blegdamsvej 17, DK--2100 Copenhagen \O, Denmark \\
$^{(2)}$Department of Physics, State University of New York, Stony Brook,
NY 11794-3800}

\date{\today}

\maketitle

\begin{abstract}
A simple lattice gas model in one dimension is constructed in which
each site can be occupied by at most one particle of any one of $D$
species.  Particles interact with a randomly drawn nearest neighbor
interaction.  This model is capable of reproducing the factorial
moments observed in high--energy scattering.  In the limit $D
\rightarrow \infty$, the factorial moments of the negative binomial
distribution are obtained naturally.
\end{abstract}


Factorial moments have provided a useful tool for the analysis of
high--energy scattering data as obtained in $e^+ e^-$ scattering
\cite{ALEPH}, $p{\bar p}$ scattering (at energies up to 900 GeV)
\cite{UA5} or the scattering of protons or heavy ions by heavy nuclei at
a projectile energy of 200 GeV/A \cite{KLM}.  One considers a range of
some variable (usually the rapidity) for which one knows the average
multiplicity, $\langle n \rangle$, and its dispersion, $\langle \Delta
n^2 \rangle$.  The data is broken into $M$ equal bins, and one
constructs the factorial moments as
\begin{equation}
F_q (M) = \left[\frac{1}{M} \sum_{i=1}^M \langle n (n-1) (n-2)
\ldots (n-q+1) \rangle_i \right] \left/ \left[\frac{1}{M} \sum_{i=1}^M
\langle n \rangle_i \right]^q \right. \ \ .
\label{1.1}
\end{equation}
Empirical factorial moments found in such high--energy scattering
experiments are well reproduced by the `negative binomial
distribution' (NB) for which
\begin{equation}
F_q^{NB} (M) = (1 + cM)(1 + 2cM) \ldots (1+[q-1]cM)
\label{1.2}
\end{equation}
with
\begin{equation}
c = \frac{\langle \Delta n^2 \rangle - \langle n \rangle}
{\langle n \rangle^2}  \ \ .
\label{1.3}
\end{equation}
The form of Eq.~(\ref{1.2}) invites the consideration of `universal'
plots of $F_q$ versus $F_2$ which permit the comparison of data from
very different processes \cite{Carruthers,Ochs}.  Such plots have been
made over the available range $1 < F_2 < 1.8$.  They reveal both a
remarkably universal behaviour and striking agreement with the
negative binomial distribution.  Giovannini and Van Hove
\cite{Giovannini} suggested that these results could be understood `in
terms of partial stimulated emission of bosons, or of a simple form of
cascade process, or (more artificially) with both mechanisms'.

The most interesting feature of Eq.~(\ref{1.2}) and the data which it
fits is the growth of the factorial moments with increasing $cM$.
This has sometimes been regarded as evidence for the presence of
fluctuations on many different scales \cite{Bialas} with the related
possibility that this may indicate the presence of critical phenomena.
Chau and Huang \cite{Chau} have offered an alternate view.  They
identified the full range of rapidity with the $N$ sites of a
one--dimensional Ising (or lattice gas) model.  They constructed the
factorial moments in this model analytically (for $N \rightarrow
\infty$) and set the two parameters of the Hamiltonian by fixing the
global values of $\langle n \rangle$ and $\langle \Delta n^2 \rangle$.
The resulting factorial moments agree through ${\cal O}(M)$ with
Eq.~(\ref{1.2}) and underestimate higher terms.  Agreement in the
constant and linear term of $F_q (M)$ is actually a trivial
consequence of (i) the fixing of $\langle n \rangle$ and $\langle
\Delta n^2 \rangle$ and (ii) the fact that the Ising model predicts
many--body correlations of a finite range \cite{footnote0}.

Our purpose here is to demonstrate the surprising result that an
elementary extension of the lattice gas model permits exact
replication of $F_q^{NB} (M)$ for all $q$ and $M$.  This extension is
simply stated: Construct a one dimensional lattice gas model (in the
$N \rightarrow \infty$ limit) where each site can either be empty or
occupied by one particle which can be of any of $D$ species.  Each
species has a chemical potential, $\mu_d$, and each pair of species
has a nearest neighbor interaction of strength $\epsilon_{dd'}$.  This
model can be solved analytically for the $q$--body correlation
functions and factorial moments using textbook techniques.  As usual,
this involves the construction and diagonalization of a matrix, ${\cal
M}$, related to the partition function for one pair of adjacent sites.
The parameters in this model can be chosen to reproduce the results of
Chau and Huang for any $D$.  Another choice of parameters will be
shown to reproduce the NB results of Eq.~(\ref{1.2}).

Solution of this model involves a $(D+1)$--dimensional real, symmetric
matrix, ${\cal M}$:
\begin{eqnarray}
{\cal M}_{00} & = & 1 \nonumber \\
{\cal M}_{0d} & = & \exp{[-\mu_d / 2]} \label{3.1} \\
{\cal M}_{dd'} & = & \exp{[-\epsilon_{dd'} -\mu_d / 2 - \mu_{d'} / 2]}
\nonumber
\end{eqnarray}
where matrix indices run from $0$ to $D$ \cite{footnote1}.  We also require
the number operator at site $i$, $n_i = \openone - T$, with $T_{dd'} =
\delta_{d0} \delta_{d'0}$.  We shall arbitrarily set all chemical potentials
equal to $\mu_o$.  In the thermodynamic limit (both $N$ and $\mu_o
\rightarrow \infty$), it is sufficient to consider the
$D$--dimensional submatrix, ${\bar {\cal M}}$, obtained by neglecting
the $0$--th row and column of ${\cal M}$.  The coupling of ${\bar {\cal M}}$
to the remaining elements of ${\cal M}$ can then be treated exactly using
first order perturbation theory.  Correlation functions and
factorial moments can be determined in terms of the diagonal form,
${\bar {\cal M}}_d$, and the related orthogonal matrix $\theta$.
(Where ${\bar {\cal M}}$ is given as $\theta {\bar {\cal M}}_d
\theta^T$.)  For example, the two--body correlation function for
microscopic sites $i$ and $i+j$ is
\begin{equation}
\langle n_i n_{i+j} \rangle = \langle n_i \rangle [ \langle n_i \rangle +
\sum_{k=1}^D a_k^2 {{\bar \lambda}_{k}}^j ]
\label{3.7}
\end{equation}
where the ${\bar \lambda}_k$ are the eigenvalues of ${\bar {\cal M}}$
\cite{footnote2} and the $a_k^2$ are normalized coefficients which
follow from the eigenvectors of ${\bar {\cal M}}$ as
\begin{equation}
{a_k} = {\cal N} \frac{1}{1 - {\bar \lambda}_k } \sum_{i=1}^D \theta_{ik}
\label{new.1}
\end{equation}
where ${\cal N}$ is chosen such that $\sum a_k^2 = 1$.  The second
factorial moment is then given as
\begin{equation}
F_2^{GLG} (M) = 1 + cM
\label{3.8}
\end{equation}
with
\begin{equation}
c = \frac{2}{\langle n\rangle} \sum_{k=1}^D a_k^2 \frac{{\bar \lambda}_k}
{1 - {\bar \lambda}_k} \ \ .
\label{3.9}
\end{equation}

After considerable algebraic effort, one finds the following expression
for the factorial moments of the generalized lattice gas model:
\begin{equation}
F_q^{GLG} (M) = \sum_{k=0}^{q-1}    \frac{q!}{k! (q-k)!}
\left[ \frac{d^k}{dz^k} [ 1 + \sum_{\mu=1}^{\infty} z^{\mu} s_{\mu} ]^{(q-k)}
\right]_{z=0} \ M^k \ \ .
\label{3.15}
\end{equation}
Here, we have introduced the definition
\begin{equation}
s_{\mu} = \frac{1}{\langle n\rangle^{\mu}} \sum_{k=1}^D
a_k^2 \left( \frac{{{\bar \lambda}}_{k}}{1-{{\bar \lambda}}_{k}}
 \right)^{\mu}
\label{3.17}
\end{equation}
which implies $s_1 = c/2$.

Eqs.~(\ref{3.15}) and (\ref{3.17}) represent the primary results of our
generalized lattice gas model.  There are two special cases of interest.
First, if
\begin{equation}
s_{\mu} = \left( \frac{c}{2} \right) ^{\mu} \ \ ,
\label{3.18}
\end{equation}
one immediately obtains the results of Chau and Huang for any $D$.
This can be realized either when the sums of Eq.~(\ref{3.17}) contain
only one term or when all the eigenvalues are equal.  The second
special case corresponds to the choice
\begin{equation}
s_{\mu} = \frac{c^{\mu}}{\mu+1} \ \ .
\label{4.1}
\end{equation}
With this choice, Eqs.~(\ref{3.15}) and (\ref{3.17}) reduce to the NB
results of Eq.~(\ref{1.2})\cite{footnote3}.

It remains to be seen if the parameters in ${\bar {\cal M}}$ can be
selected so that the constraints of Eq.~(\ref{4.1}) are satisfied.
The following simple prescription works: Set all the off--diagonal
elements of ${\bar {\cal M}}$ equal to zero.  Make a draw of $D$
random numbers, $x_d$, from the interval $[0,1]$.  Set ${\bar {\cal
M}}_{dd}$ equal to $x_d L$.  For each draw, choose $L$ such that $s_1
= c/2$.  (This sets the dispersion to its empirical value for each
draw.)  The physical content of this prescription is clear.
Identical particles experience a nearest neighbor interaction which
ranges from $\approx - \mu_o$ to $+\infty$.  Inequivalent particles
experience a nearest neighbor interaction in the range $-\mu_o \ll
\epsilon_{dd'} \leq +\infty$.  More `democratic' schemes can also be
constructed.  They share the feature that ${\bar {\cal M}}$ is sparse
with roughly $D$ (randomly selected) elements non--zero
\cite{JWB}.

Our prescription meets the remaining conditions of Eq.~(\ref{4.1})
with increasing accuracy as $D \rightarrow \infty$.  We shall
illustrate this with a numerical study of the first seven factorial
moments.  We take $\langle n\rangle = 20$ and $\langle\Delta n^2
\rangle = 110$ as would be appropriate for the description of $p{\bar
p}$ scattering at 200 GeV.  For these data, $c \langle n \rangle =
4.5$.  (Qualitatively similar results are obtained for 900 GeV $p{\bar
p}$ scattering.)  We consider the ratios
\begin{equation}
r_q = \frac{(q+1) \sum_{k=1}^D a_k^2 \left( \frac{{{\bar \lambda}}_{k}}
{1-{{\bar \lambda}}_{k}}\right)^{q}}{\left[c \langle n\rangle \right]^q} \ \ .
\label{4.9}
\end{equation}
These ratios should be $r_q = 1$ to reproduce the negative binomial
distribution.  Our constraint ensures $r_1 = 1$.  For each value of
$D$, we have drawn $10^5$ matrices according to the prescription
above.  In Table~\ref{table1} we report the `ensemble average over
theories', $\langle\langle r_q \rangle\rangle$, and its dispersion for
$2 \le q \le 6$ \cite{footnote4} as obtained for $D=1$, $2$, $4$, \ldots,
$512$.  Since there are no parameters to adjust, this is an extremely
stringent test of our prescription.  It succeeds.

Several comments are in order.  For fixed $q$, the value of
$\langle\langle r_q \rangle\rangle$ approaches $1$ like $1/D$ as $D
\rightarrow \infty$.  The dispersion also vanishes (like
$1/\sqrt{D}$).  Thus, as $D$ becomes large, our simple prescription
converges to the results of the NB for any fixed $q$.  For fixed $D$,
the error in $\langle\langle r_q \rangle\rangle$ and its dispersion
grow as $q \rightarrow \infty$.  The value of $D = 16$ results in
sufficiently small errors and dispersions that randomly drawn dynamics
have a high probability of reproducing the empirical factorial moments
for $p{\bar p}$ scattering at 200 GeV within existing experimental
uncertainties.  This value of $D=16$ is also sufficient to provide a
quantitative description of factorial moments for 900 GeV $p{\bar p}$
scattering and, indeed, of all other high--energy scattering
experiments for which factorial moments are known.  Our point here is
that there exists at least one simple prescription for satisfying
Eq.~(\ref{4.1}).  Other more efficient prescriptions may well exist.

We have shown that a simple one--dimensional lattice gas model with
$D$ species and randomly drawn nearest neighbor interactions between
equivalent species can reproduce the factorial moments of the negative
binomial distribution as $D$ becomes large.  Given the limited range
of $cM$ covered by current experimental data, a remarkably small
number of species is sufficient to provide a quantitative description
of the empirical factorial moments.  This offers some understanding
for both the success of cascade calculations and the anecdotal
observation that the results of such calculations are often
surprisingly insensitive to the details of the model.  In the present
picture, any of our randomly drawn theories would also be likely to
succeed (at least at the level of the factorial moments).

The empirical observation that factorial moments grow like powers of
$cM$ in $p{\bar p}$, $e^+ e^-$ and relativistic heavy ion collisions
(the phenomenon of intermittency) has often be taken as evidence of
the existence of fluctuations on `all length scales'.  As such, it is
sometimes seen as an indicator of the presence of a non--equilibrium,
critical phenomenon.  While we do not deny the possibility that
intermittency {\em can\/} be a signature of critical phenomena, we
have shown that a simple but highly heterogeneous (equilibrium) system
can also lead to intermittency.  In short, intermittency is not a
unique signature of critical phenomena.  Given the small number of
species required by our model to fit the factorial moments obtained in
high--energy $p{\bar p}$ scattering, a critical phenomenon does not
even appear to be the `most plausible' cause of intermittency.

Two of us (ADJ and TW) would like to acknowledge the hospitality of
NORDITA.  This work was partially supported by the U.S. Department of
Energy under grant no. \mbox{DE-FG02-88ER 40388}.

\begin{table}
\caption{The ensemble average and dispersion of $r_q$ as defined in
Eq.~(\protect\ref{4.9}) for $2 \le q \le 6$ and various values of the
number of particle species, $D$. The average was taken over $10^5$
randomly drawn theories with $c \langle n \rangle = 4.5$ for each draw.
The case $D=1$ corresponds to the ordinary lattice gas (Ising) model
\protect\cite{Chau} and has no dispersion.}
\label{table1}
\begin{tabular}{rr@{${}\pm{}$}lr@{${}\pm{}$}lr@{${}\pm{}$}l
                 r@{${}\pm{}$}lr@{${}\pm{}$}l}
 \multicolumn{1}{c}{$D$} &
 \multicolumn{2}{c}{$\langle\langle r_2 \rangle\rangle$} &
 \multicolumn{2}{c}{$\langle\langle r_3 \rangle\rangle$} &
 \multicolumn{2}{c}{$\langle\langle r_4 \rangle\rangle$} &
 \multicolumn{2}{c}{$\langle\langle r_5 \rangle\rangle$} &
 \multicolumn{2}{c}{$\langle\langle r_6 \rangle\rangle$} \\ \tableline
   1 & \multicolumn{2}{c}{0.75} & \multicolumn{2}{c}{0.5} &
       \multicolumn{2}{c}{0.3125} & \multicolumn{2}{c}{0.1875} &
       \multicolumn{2}{c}{0.109375} \\
   2 & 0.812 & 0.023 & 0.598 & 0.035 & 0.417 & 0.037 & 0.280 & 0.033 &
       0.184 & 0.027 \\
   4 & 0.881 & 0.046 & 0.723 & 0.078 & 0.569 & 0.091 & 0.435 & 0.091 &
       0.326 & 0.083 \\
   8 & 0.941 & 0.067 & 0.851 & 0.132 & 0.751 & 0.179 & 0.652 & 0.207 &
       0.558 & 0.218 \\
  16 & 0.981 & 0.076 & 0.951 & 0.173 & 0.916 & 0.270 & 0.879 & 0.360 &
       0.842 & 0.441 \\
  32 & 0.998 & 0.069 & 0.999 & 0.170 & 1.005 & 0.295 & 1.017 & 0.442 &
       1.039 & 0.612 \\
  64 & 1.002 & 0.050 & 1.007 & 0.128 & 1.019 & 0.229 & 1.040 & 0.355 &
       1.071 & 0.515 \\
 128 & 1.001 & 0.035 & 1.005 & 0.088 & 1.013 & 0.154 & 1.026 & 0.233 &
       1.044 & 0.327 \\
 256 & 1.001 & 0.024 & 1.003 & 0.061 & 1.007 & 0.105 & 1.013 & 0.155 &
       1.023 & 0.210 \\
 512 & 1.000 & 0.017 & 1.002 & 0.042 & 1.004 & 0.072 & 1.007 & 0.106 &
       1.012 & 0.142 \\
\end{tabular}
\end{table}


\begin{references}
\bibitem{footnote0} This second feature is common to all one--dimensional
  models.
\bibitem{footnote1} Without loss of generality, we have set the inverse
  temperature, $\beta$, equal to $1$.
\bibitem{footnote2} We consider only the physically interesting
  case where $|{{\bar \lambda}_{k}}| < 1$ for all $k$.
\bibitem{footnote4} Only five factorial moments are currently available
  experimentally.  Therefore, only $r_2$ to $r_4$ are relevant for
  comparison with data.
\end{references}
\end{document}